\documentclass{cupconf}
\usepackage{epsfig}

  \checkfont{eurm10}
  \iffontfound
    \IfFileExists{upmath.sty}
      {\typeout{^^JFound AMS Euler Roman fonts on the system,
                   using the 'upmath' package.^^J}%
       \usepackage{upmath}}
      {\typeout{^^JFound AMS Euler Roman fonts on the system, but you
                   dont seem to have the}%
       \typeout{'upmath' package installed. cupconf.cls can take advantage
                 of these fonts,^^Jif you use 'upmath' package.^^J}%
      }
  \else
  \fi

  \checkfont{msam10}
  \iffontfound
    \IfFileExists{amssymb.sty}
      {\typeout{^^JFound AMS Symbol fonts on the system, using the
                'amssymb' package.^^J}%
       \usepackage{amssymb}%
         
       \let\ge=\geqslant  
      }{}
  \fi

  \IfFileExists{amsbsy.sty}
    {\typeout{^^JFound the 'amsbsy' package on the system, using it.^^J}%
     \usepackage{amsbsy}}
    {\providecommand\boldsymbol[1]{\mbox{\boldmath $##1$}}}

\newsavebox{\astrutbox}
\sbox{\astrutbox}{\rule[-5pt]{0pt}{20pt}}

\newcommand\etal{\mbox{\textit{et al.}}}

\title[Hot Baryons in Supercluster Filaments]{{\it \small \flushleft STScI May Symposium, 3--6 May, 2004, Baltimore, MD, USA\\Planets To Cosmology:  Essential Science In Hubble's Final Years\\
M. Livio (ed.)\\\mbox{ }\\} Hot Baryons in Supercluster Filaments}

\author[E. D. Miller {\it et al.\/}]%
{Eric D. Miller, Renato A. Dupke, and Joel N. Bregman}

\affiliation{Department of Astronomy, University of Michigan,
Ann Arbor, MI 48109, USA}

\begin{document}

%%%%%%%%%%%%%%%%%%%%%%%%
%%%%%%%%%%%%%%%%%%%%%%%%
%%%%%%%%%%%%%%%%%%%%%%%%

\newcommand\eex[1]{\mbox{$\times 10^{#1}$}}              % #1 x 10^#2
\newcommand\eez[1]{\mbox{$10^{#1}$}}                     % just 10^#2
\newcommand\ion[2]{\mbox{#1$\,${\small\rmfamily{#2}}}}   % \ion{Fe}{2} => FeII
\newcommand\nodata{ ~$\cdots$~ }% 
\newcommand\nd{\nodata}
\def\ovi{\ion{O}{VI}}
\def\hi{\ion{H}{I}}
\def\lya{\mbox{Ly$\alpha$}}
\def\lyb{\mbox{Ly$\beta$}}
\def\lyg{\mbox{Ly$\gamma$}}
\def\zagn{\mbox{$z_{\rm AGN}$}}
\def\zsc{\mbox{$z_{\rm SC}$}}
\def\kps{\mbox{${\rm km~s^{-1}}$}}
\def\lam{\mbox{$\lambda$}}
\def\ew{\mbox{$W_{\lambda}$}}

\def\jnl@style{\it}
%commente par Seb
\def\aaref@jnl#1{{\jnl@style#1}}
%ref remplace par aaref pour eviter conflit...

\def\aaref@jnl#1{{\jnl@style#1}}

\def\aj{\aaref@jnl{AJ}}                   % Astronomical Journal
\def\araa{\aaref@jnl{ARA\&A}}             % Annual Review of Astron and Astrophys
\def\apj{\aaref@jnl{ApJ}}                 % Astrophysical Journal
\def\apjl{\aaref@jnl{ApJ}}                % Astrophysical Journal, Letters
\def\apjs{\aaref@jnl{ApJS}}               % Astrophysical Journal, Supplement
\def\ao{\aaref@jnl{Appl.~Opt.}}           % Applied Optics
\def\apss{\aaref@jnl{Ap\&SS}}             % Astrophysics and Space Science
\def\aap{\aaref@jnl{A\&A}}                % Astronomy and Astrophysics
\def\aapr{\aaref@jnl{A\&A~Rev.}}          % Astronomy and Astrophysics Reviews
\def\aaps{\aaref@jnl{A\&AS}}              % Astronomy and Astrophysics, Supplement
\def\azh{\aaref@jnl{AZh}}                 % Astronomicheskii Zhurnal
\def\baas{\aaref@jnl{BAAS}}               % Bulletin of the AAS
\def\jrasc{\aaref@jnl{JRASC}}             % Journal of the RAS of Canada
\def\memras{\aaref@jnl{MmRAS}}            % Memoirs of the RAS
\def\mnras{\aaref@jnl{MNRAS}}             % Monthly Notices of the RAS
\def\pra{\aaref@jnl{Phys.~Rev.~A}}        % Physical Review A: General Physics
\def\prb{\aaref@jnl{Phys.~Rev.~B}}        % Physical Review B: Solid State
\def\prc{\aaref@jnl{Phys.~Rev.~C}}        % Physical Review C
\def\prd{\aaref@jnl{Phys.~Rev.~D}}        % Physical Review D
\def\pre{\aaref@jnl{Phys.~Rev.~E}}        % Physical Review E
\def\prl{\aaref@jnl{Phys.~Rev.~Lett.}}    % Physical Review Letters
\def\pasp{\aaref@jnl{PASP}}               % Publications of the ASP
\def\pasj{\aaref@jnl{PASJ}}               % Publications of the ASJ
\def\pasa{\aaref@jnl{PASA}}               % Publications of the AS of Austral.
\def\qjras{\aaref@jnl{QJRAS}}             % Quarterly Journal of the RAS
\def\skytel{\aaref@jnl{S\&T}}             % Sky and Telescope
\def\solphys{\aaref@jnl{Sol.~Phys.}}      % Solar Physics
\def\sovast{\aaref@jnl{Soviet~Ast.}}      % Soviet Astronomy
\def\ssr{\aaref@jnl{Space~Sci.~Rev.}}     % Space Science Reviews
\def\zap{\aaref@jnl{ZAp}}                 % Zeitschrift fuer Astrophysik
\def\nat{\aaref@jnl{Nature}}              % Nature
\def\iaucirc{\aaref@jnl{IAU~Circ.}}       % IAU Cirulars
\def\aplett{\aaref@jnl{Astrophys.~Lett.}} % Astrophysics Letters
\def\apspr{\aaref@jnl{Astrophys.~Space~Phys.~Res.}}
                % Astrophysics Space Physics Research
\def\bain{\aaref@jnl{Bull.~Astron.~Inst.~Netherlands}} 
                % Bulletin Astronomical Institute of the Netherlands
\def\fcp{\aaref@jnl{Fund.~Cosmic~Phys.}}  % Fundamental Cosmic Physics
\def\gca{\aaref@jnl{Geochim.~Cosmochim.~Acta}}   % Geochimica Cosmochimica Acta
\def\grl{\aaref@jnl{Geophys.~Res.~Lett.}} % Geophysics Research Letters
\def\jcp{\aaref@jnl{J.~Chem.~Phys.}}      % Journal of Chemical Physics
\def\jgr{\aaref@jnl{J.~Geophys.~Res.}}    % Journal of Geophysics Research
\def\jqsrt{\aaref@jnl{J.~Quant.~Spec.~Radiat.~Transf.}}
                % Journal of Quantitiative Spectroscopy and Radiative Transfer
\def\memsai{\aaref@jnl{Mem.~Soc.~Astron.~Italiana}}
                % Mem. Societa Astronomica Italiana
\def\nphysa{\aaref@jnl{Nucl.~Phys.~A}}   % Nuclear Physics A
\def\physrep{\aaref@jnl{Phys.~Rep.}}   % Physics Reports
\def\physscr{\aaref@jnl{Phys.~Scr}}   % Physica Scripta
\def\planss{\aaref@jnl{Planet.~Space~Sci.}}   % Planetary Space Science
\def\procspie{\aaref@jnl{Proc.~SPIE}}   % Proceedings of the SPIE

\def\crasp{\aaref@jnl{C.~R.~Acad.~Sci.~Paris}}   % Muller Oort Raimond journ

\let\astap=\aap
\let\apjlett=\apjl
\let\apjsupp=\apjs
\let\applopt=\ao

\maketitle

\begin{abstract}
Most of the baryons in the local universe are ``missing'' in that they are
not in galaxies or in the previously detected gaseous phases. These missing
baryons are predicted to be in a moderately hot phase, 10$^5$--10$^7$ K,
largely in the form of giant cosmic filaments that connect the denser
virialized clusters and groups of galaxies. Models show that the highest
covering fraction of such filaments occurs in superclusters. To determine
whether such filaments exist, we have begun a project to search for UV
absorption against AGNs projected behind possible supercluster filaments.
Using data from the HST and FUSE archives along with new observations, we
have detected UV absorption within about 1300 km/s of seven supercluster
sightlines out of a sample of eight.  The likelihood of such detections
being generated by chance is less than 10$^{-4}$.
\end{abstract}

\firstsection % if your document starts with a section,
              % remove some space above using this command.
\section{Introduction}\label{sec:intro}

A census of baryons in the local universe indicates that the majority of
this normal matter is undetected, or ``missing''.  At high redshifts ($z
\sim 3$), big-bang nucleosynthesis models and QSO absorption line
observations indicate a baryon mass fraction of $\Omega_b \sim 0.04$ (e.g.,
Fukugita, Hogan \& Peebles 1998).  The stars and gas detected in local
galaxies account for only 20\% of this ($\Omega_b \sim 0.008$).
The absence of a local Ly$\alpha$ forest indicates that these baryons are
likely in a hot ($T > \eez{5}$ K), diffuse medium which has heretofore
remained undetectable (e.g., Fukugita, Hogan \& Peebles 1998; Cen \&
Ostriker 1999a; Dav\'{e} \etal\ 2001).

A reservoir of hot gas would be organized similarly to the collapsed
structure, i.e. into a web of filaments like those seen in structure
formation simulations (e.g. Evrard \etal\ 2002).  Gas is shock-heated as it
collapses into structures, and at the nodes of this web are found the rich
clusters of galaxy, which contain large concentrations of hot gas seen in
X-ray emission ($T \sim \eez{7}$--$\eez{8}$).  Galaxy groups are found
along the filaments connecting nodes, and while several groups are detected
in diffuse X-ray emission ($T \sim \eez{7}$ K), the extent of this gaseous
component is not well-constrained.  The filaments themselves contain most
of the volume of collapsed structure, and the temperature in these regions
is thought to be lower than that of the denser clusters and groups, or $T
\sim \eez{5}$--$\eez{7}$ K (e.g., Cen \etal\ 1995).  

Recent results have shown possible absorption by this hot cosmic web of
baryons against background point sources (Tripp \& Savage 2000; Savage
\etal\ 2002; Richter \etal\ 2004).  We have begun a project to (1) identify
lines connecting clusters within superclusters and (2) search for
absorption against AGNs projected behind these likely filament locations.
Results from three AGNs probing four superclusters have been given in
detail by Bregman, Dupke \& Miller (2004). Here we summarize that study and
present preliminary results for four additional AGN/supercluster
sightlines.  This project makes use of both archival and proprietary HST
data, using all generations of spectrographs, from FOS to GHRS to STIS.  As
we primarily search for \lya\ \lam1216 absorption at low redshift, this
project requires space-based UV spectroscopy and would be impossible
without the sensitivity and wavelength coverage of HST.

\section{Identifying Filaments and Background AGNs}

Filaments themselves are not visible in emission to current instruments, so
knowing where to look requires predicting their locations.  Structure
formation simulations indicate that filaments should connect clusters in
approximately straight lines (e.g. Evrard \etal\ 2002), with the width of
the filaments on the order of the virial radius of a typical cluster
($3h^{-1}$ Mpc; Cen \& Ostriker 1999b).  The supercluster catalog of 
Einasto \etal\ (1997), which contains 220 superclusters at $z < 0.12$,
allows us to visually identify filaments connecting clusters in space and
redshift and cross-correlate these locations with AGNs from the
V\'{e}ron-Cetty \& V\'{e}ron (2001) catalog.
Observation with HST and FUSE necessitates selection of bright ($V < 16$)
AGNs, which are typically nearby Seyfert galaxies at $z < 0.2$, and it
requires low Galactic extinction (from Schlegel, Finkbeiner \& Davis 1998).
The background AGNs must be well-separated in redshift from the intervening
superclusters ($\zagn \gg \zsc$), and these combined requirements
preferentially select nearby superclusters.  The larger solid angle
subtended by these nearer systems also increases the likelihood of finding a
suitable background AGN within $\sim$ 3 Mpc of the predicted filament line.

\begin{figure}
\begin{minipage}[c]{\linewidth}
\begin{minipage}{.5\linewidth}
\epsfig{file=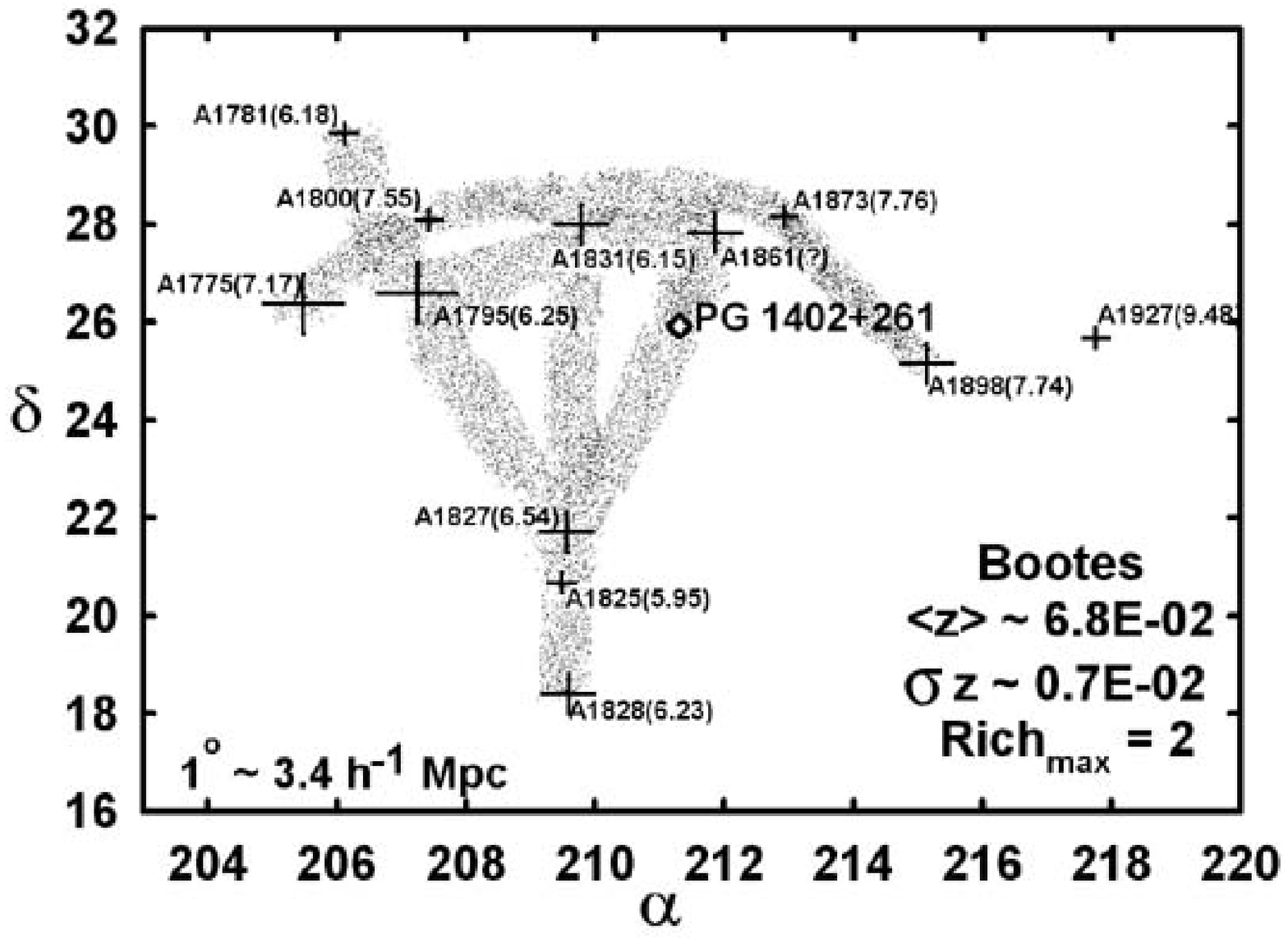,width=\linewidth}
\end{minipage}\hfill
\begin{minipage}{.5\linewidth}
\epsfig{file=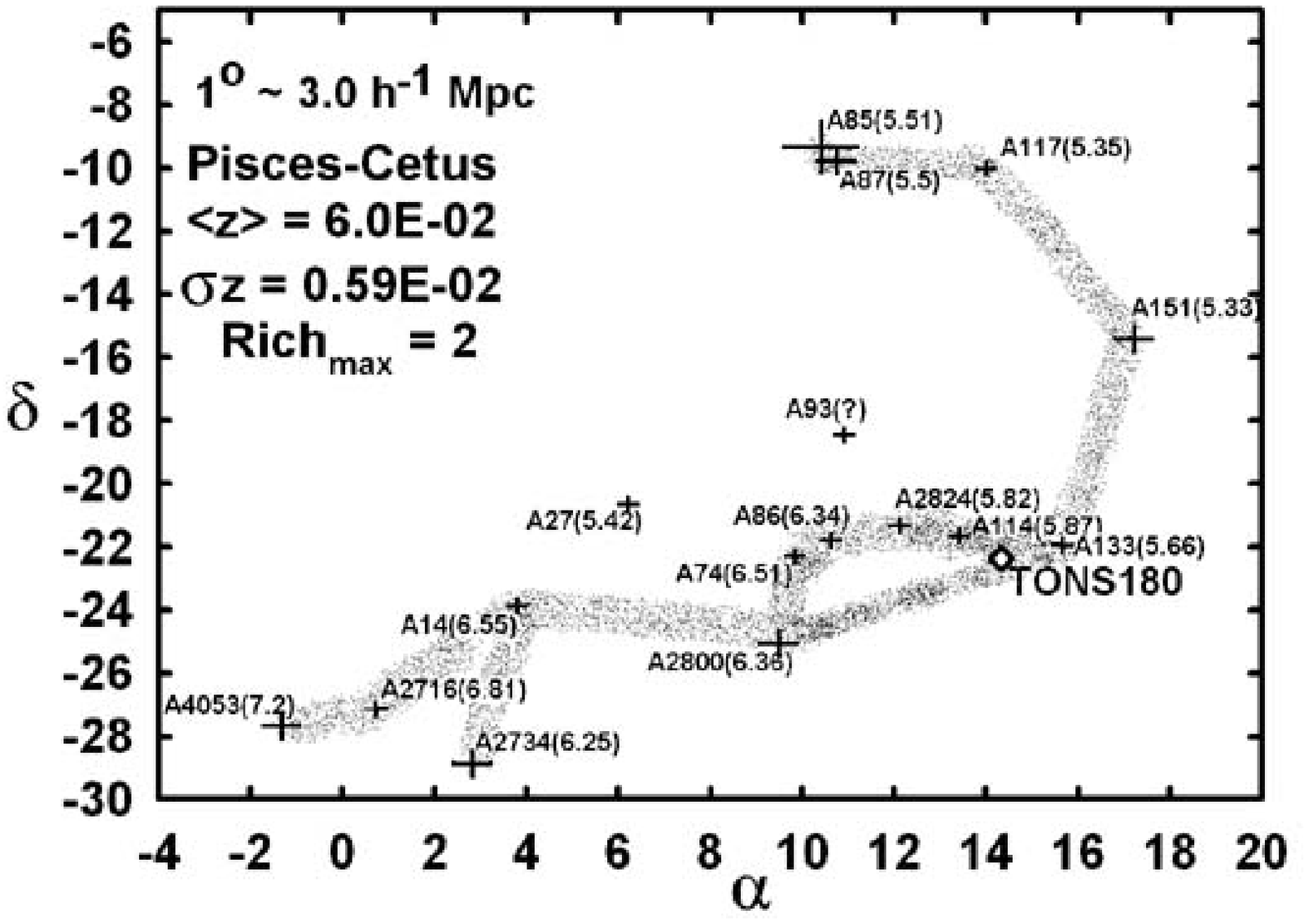,width=\linewidth}
\end{minipage}
\end{minipage}
\begin{minipage}[b]{\linewidth}
\begin{minipage}{.5\linewidth}
\epsfig{file=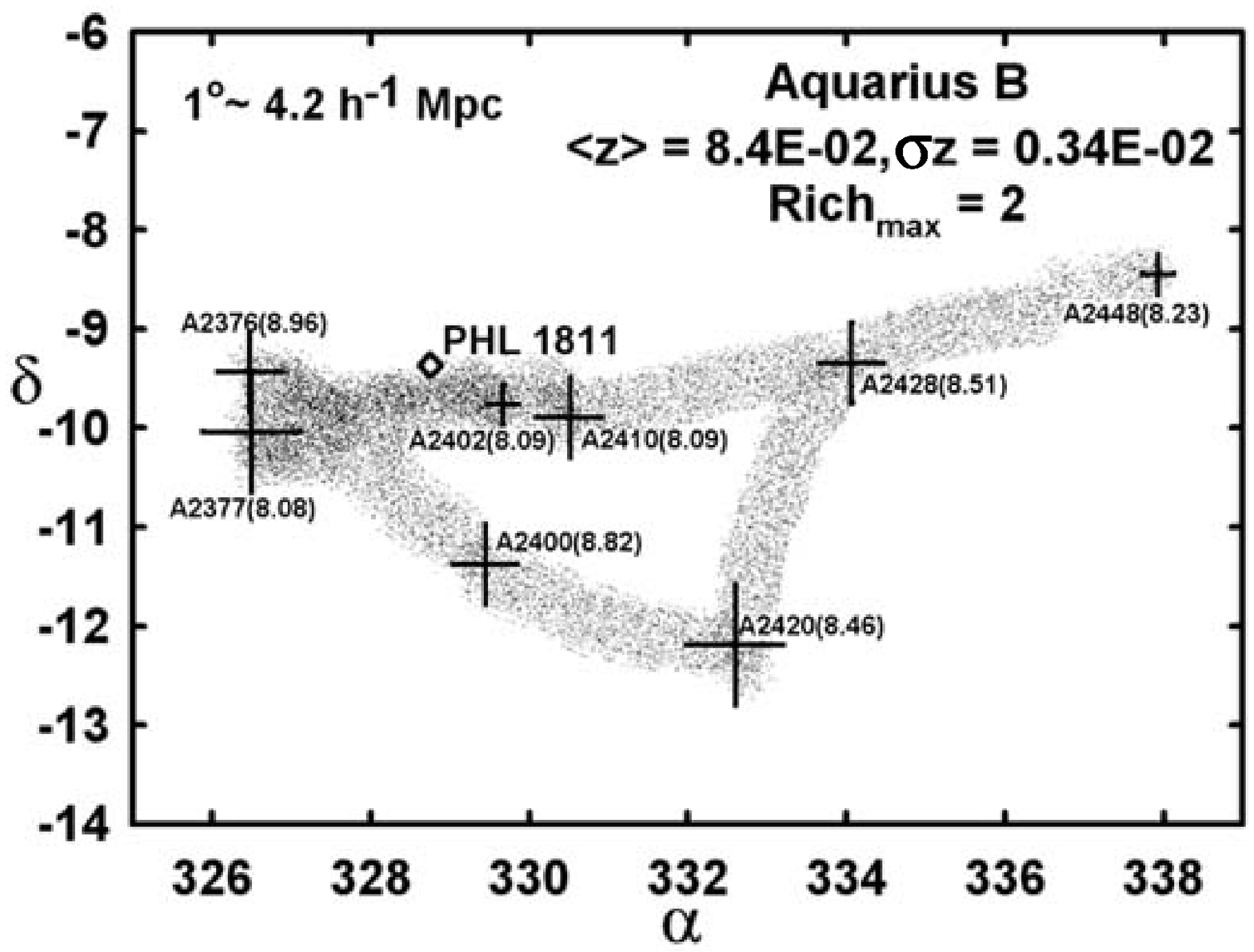,width=\linewidth}
\end{minipage}\hfill
\begin{minipage}{.5\linewidth}
\epsfig{file=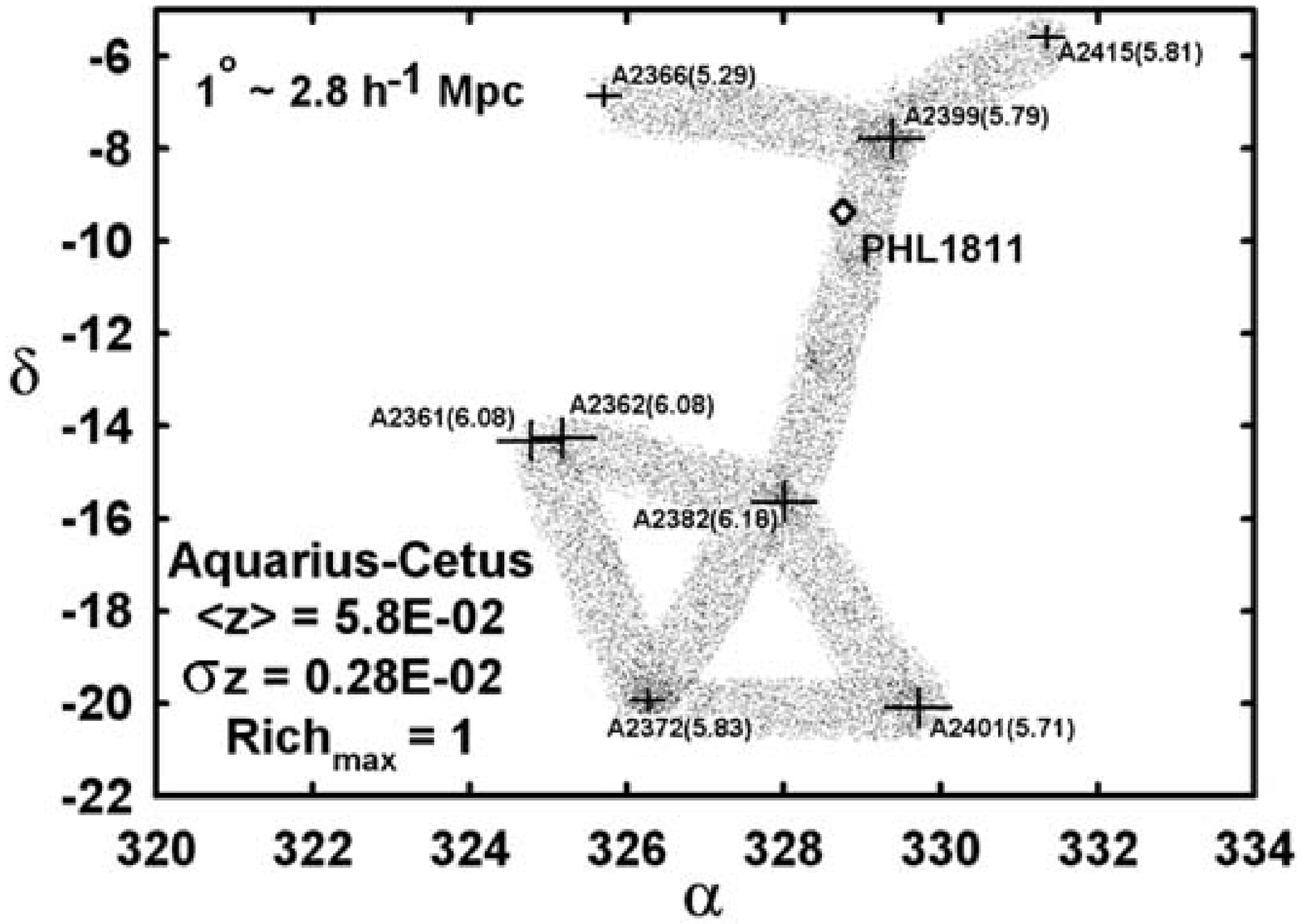,width=\linewidth}
\end{minipage}
\end{minipage}
\caption{Spatial configuration of background AGNs (open diamonds) projected
onto four superclusters. The crosses are galaxy clusters (many with Abell
names), where the size of the cross indicates the relative richness. The
linear grey stripes are connections between galaxy clusters that are
possible filaments projected near the background AGN, based upon
three-dimensional proximity and cluster richness; these filaments are 3 Mpc
wide. We also indicate the redshift of the clusters in parenthesis (in
units of $10^{-2}$), the average redshift of the supercluster, and the
standard deviation of cluster redshifts in the supercluster.
\label{fig:scmaps}}
\end{figure}

We have identified 11 supercluster sightlines probed by 10 AGNs (PHL 1811
lies behind two distinct superclusters), and these are listed in Table
\ref{tab:tab1}.  Maps of four superclusters are shown in Figure
\ref{fig:scmaps} to demonstrate the predicted filament structure and
projected locations of background AGNs.  At publication time, seven AGNs
(probing eight supercluster sightlines) have been observed by HST, and four
AGNs (probing five superclusters) have been observed by FUSE; these results
include a combination of archived and proprietary data.  The object Ton
S180 fails our $\zagn \gg \zsc$ criterion, and we treat it as a special
case in the following discussion.  Data for four of the AGNs have been
discussed in the literature: PHL 1811, studied extensively by Jenkins 
\etal\ (2003); PG 1402+261, studied by Bechtold \etal\ (2002) and Wakker
\etal\ (2003); H1821+643, observed by Tripp, Lu \& Savage (1998) and
Oegerle \etal\ (2000); and Ton S180, reported on by a number of authors
(Shull \etal\ 2000; Turner \etal\ 2002; Wakker \etal\ 2003; Penton,
Stocke \& Shull 2004).  We have incorporated their results and our own
analysis of the data here.  The remainder of the AGNs were observed by us
for this project with the exception of KAZ 102, which was obtained from the
HST archive.

\begin{table}
\begin{minipage}{\linewidth}
\begin{center}
\begin{tabular}{llccccrrr}
AGN             & supercluster(s)      & $z_{\rm AGN}$ & $z_{\rm SC}$ & HST       & FUSE      & $W_{\rm Ly}$ & $W_{\rm OVI}$ & $T_{\rm max}$ \\
                &                      &       &       &           &           & (m\AA)    &  (m\AA)    & ($10^6$ K) \\[3pt]
PHL1811         & Aquarius B           & 0.192 & 0.084 & $\bullet$ & $\bullet$ & 300     &  50      & 0.2 \\
                & Aquarius-Cetus       & 0.192 & 0.056 & $\bullet$ & $\bullet$ & 600     &  $<$250  & 1.0 \\
PG1402+261      & Bootes               & 0.164 & 0.068 & $\bullet$ & $\bullet$ & 300     &  $<$100  & 1.6 \\
Ton S180        & Pisces-Cetus         & 0.062 & 0.060 & $\bullet$ & $\bullet$ & 100     &  150     & 0.1 \\
KAZ102          & North Ecliptic Pole  & 0.136 & 0.087 & $\bullet$ & \nd       & 400     &  \nd     & 6.4 \\
H1821+643       & North Ecliptic Pole  & 0.297 & 0.087 & $\bullet$ & $\bullet$ & 50      &  $<$150  & 0.4 \\
WGAJ2153        & Aquarius-Cetus       & 0.078 & 0.056 & $\bullet$ & \nd       & 400     &  \nd     & 3.6   \\
RXSJ01004-5113  & Phoenix              & 0.062 & 0.027 & $\bullet$ & \nd       & $<$150  &  \nd     & \nd \\
RXSJ01149-4224  & Phoenix              & 0.124 & 0.027 & $\circ$   & $\circ$   & \nd     & \nd      & \nd \\
HE0348-5353     & Horologium-Reticulum & 0.130 & 0.064 & $\circ$   & \nd       & \nd     & \nd      & \nd \\
TEX1601+160     & Hercules             & 0.109 & 0.035 & $\circ$   & \nd       & \nd     & \nd      & \nd \\
Ton 730         & Bootes               & 0.087 & 0.065 & $\circ$   & $\circ$   & \nd     & \nd      & \nd \\
%\footnote{The PHL1811 sightline probes two superclusters.}
%\footnote{For Ton S180, $z_{\rm AGN} \sim z_{\rm SC}$, therefore the Ly and \ovi systems might be due to AGN self-absorption.}
\end{tabular}
\end{center}
\end{minipage}
\caption{Summary of AGN sightlines and detected absorption by intervening
supercluster filaments.  Filled circles indicate that data have been
obtained for the given instrument while open circles indicate targets that
have been approved.  The last column shows the upper limit on the
temperature of the absorbing medium, using the width of the narrowest
absorption feature as an upper limit to thermal line broadening.}
\label{tab:tab1}
\end{table}

\section{Absorption Line Results}

The HST and FUSE spectra together probe a number of high- and
low-ionization species, including the strong lines of \lya\ \lam1216, \lyb\
\lam1026, \ovi\ \lam\lam1032,1038, \ion{C}{II} \lam1335, and \ion{C}{III}
\lam977.  At the expected filament temperatures of about 3\eex{5} K, the
low-ionization metal lines will be absent, but the Ly series of \hi\ will
still be present.  The relative strengths of the Ly and \ovi\ absorption
depend strongly on such parameters as temperature and metalicity.  We
expect individual filaments to have a velocity dispersion no greater than
that of a typical cluster ($\sim 1300$ \kps), centered on the redshift of
the nearest clusters, so we have searched for absorption from all available
species within $\pm 1300$ \kps\ of this average velocity. 

\begin{figure}
\begin{minipage}[t]{\linewidth}
\centering{\epsfig{file=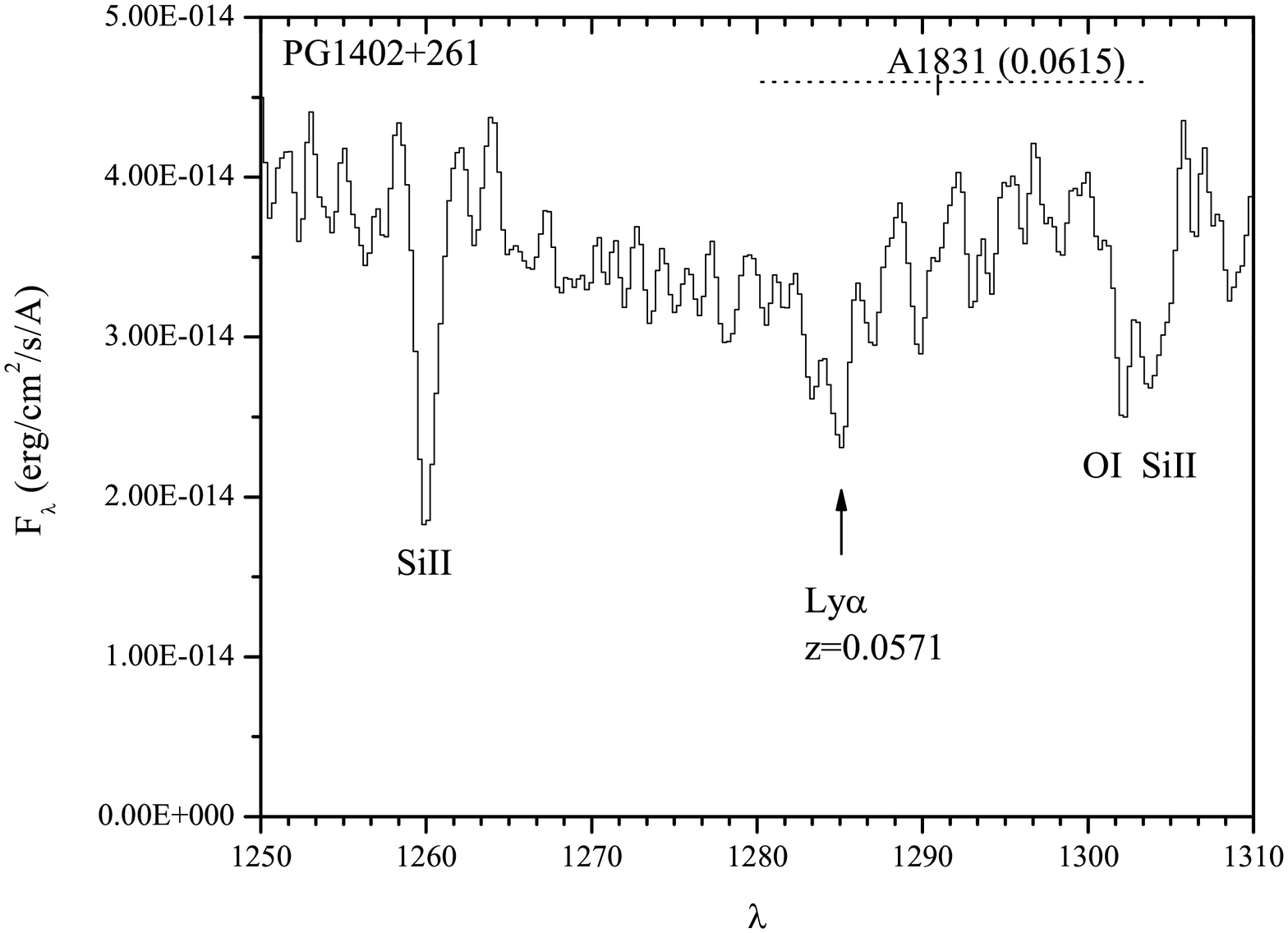,width=0.85\linewidth}}
\caption{HST/FOS spectrum of PG 1402+261 showing absorption due to Galactic
metal lines and a \lya\ system near the Bootes supercluster redshift.  The
dotted line shows the expected region of absorption, within $\pm 1300$
\kps\ of the supercluster redshift.
\label{fig:spec1}}
\end{minipage}
\begin{minipage}[b]{\linewidth}
\centering{\epsfig{file=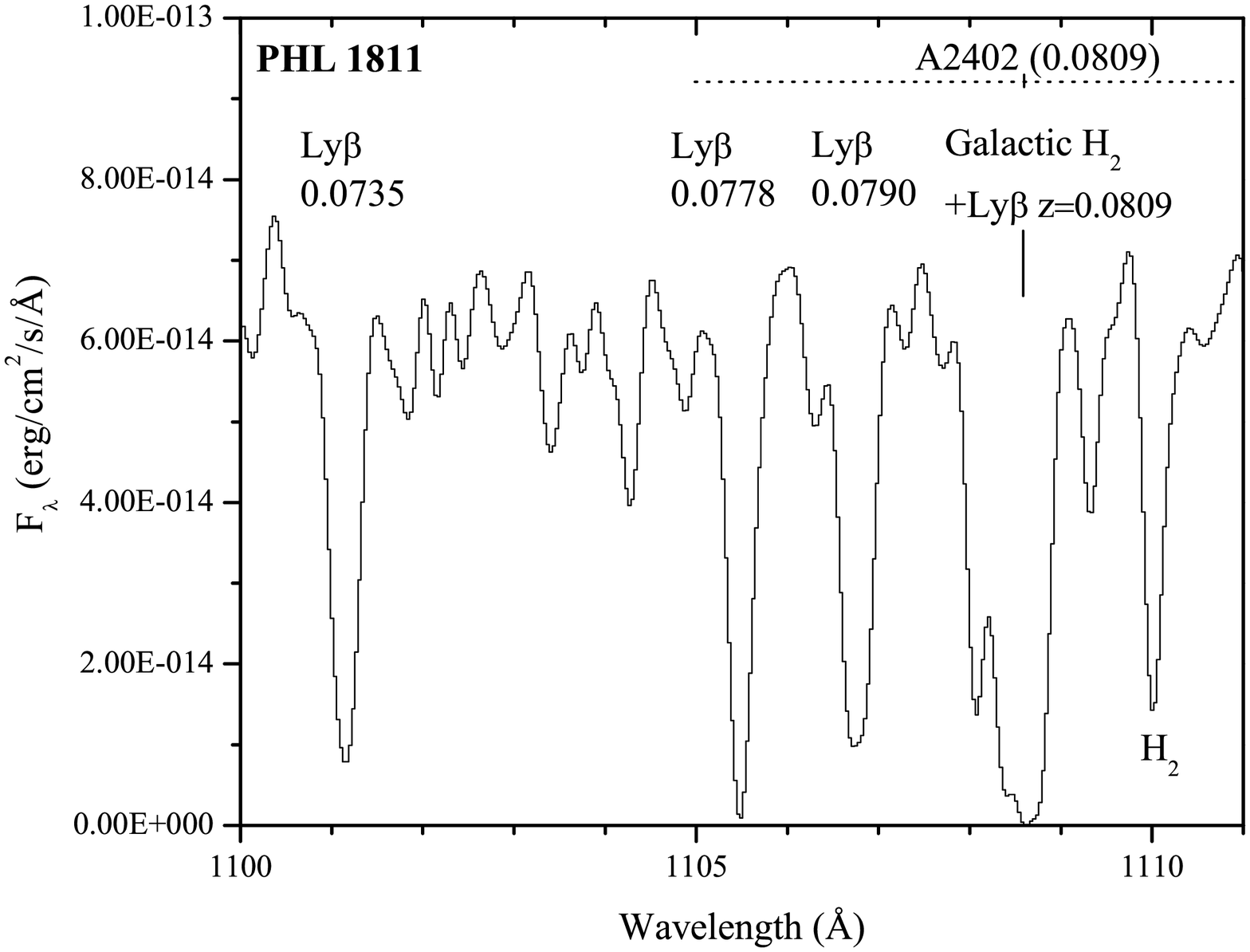,width=0.85\linewidth}}
\caption{FUSE spectrum of PHL 1811 showing four \lyb\ systems identified by
Jenkins \etal\ (2003), three of which are near the redshift of the Aquarius
B supercluster.  Absorption from Galactic H$_2$ is also evident.
\label{fig:spec2}}
\end{minipage}
\end{figure}

Absorption from \lya\ is detected in seven of the eight superclusters
observed with HST, and \ovi\ absorption is seen in two of the five 
sightlines probed with FUSE.  The \lya\ and \ovi\ \lam1032 line strengths (or
upper limits) are listed in Table \ref{tab:tab1}.  Line widths were
measured by fitting a Gaussian to the line profile, and from this we
obtained a maximum temperature corresponding to the maximum allowed thermal
line broadening.  As is shown in Table \ref{tab:tab1}, these temperatures
range from \eez{5}--\eez{6} K, however it must be stressed that these are
upper limits and it is possible the lines are composed of multiple, blended
components.  Two sample spectra are presented in Figures \ref{fig:spec1}
and \ref{fig:spec2}, showing \lya\ absorption near the redshift of the
Bootes supercluster (against PG 1402+261) and multiple \lyb\ systems near
the redshift of the Aquarius B supercluster (against PHL 1811).

%\begin{figure}
%\begin{minipage}{.5\linewidth}
%\epsfig{file=miller.f2.eps,width=\linewidth}
%\end{minipage}\hfill
%\begin{minipage}{.5\linewidth}
%\epsfig{file=miller.f3.eps,width=\linewidth}
%\end{minipage}
%\caption{{\it Left:\/} HST/FOS spectrum showing absorption due to Galactic
%metal lines and a \lya\ system near the Bootes supercluster redshift.  {\it
%Right:\/} FUSE spectrum showing four \lyb\ systems identified by Jenkins
%\etal\ (2003), three of which are near the redshift of the Aquarius B
%supercluster.  Absorption from Galactic H$_2$ is also evident.
%\label{fig:spectra}}
%\end{figure}

\section{Galaxy Halo Absorption?}

It has been been suggested that low-$z$ \lya\ absorbers are associated with
galaxies, and specifically that all strong \lya\ systems (with $\ew \ge
240$ m\AA) arise from absorption within extended gaseous halos of galaxies
near the line of sight (Lanzetta \etal\ 1995; Chen \etal\ 1998; Chen \etal\ 
2001).  Under this model, and given the \ew\ measurements shown in Table
\ref{tab:tab1}, it is possible some of the observed absorption is due to
intervening galaxies and not filaments.

One questionable case is the absorption against PHL 1811 near the redshift
of the Aquarius B supercluster.  Four \lyb\ systems are seen by Jenkins
\etal\ (2003) at redshifts of 0.0735, 0.0778, 0.0790 and 0.0809 (see Figure
\ref{fig:spec2}), with \ovi\ observed at two of these redshifts, 0.0778 and
0.0809.  Jenkins \etal\ (2003) identify an $L^*$ galaxy at the redshift of
the 0.0809 system (a Lyman limit system), projected 34 kpc away.  The other
three systems are shifted 580--2250 \kps\ in velocity and unlikely to be
associated with gravitationally bound material in this galaxy.  The authors
suggest this material is due to tidal interactions, but the lack of tidal
features in the LLS galaxy make this unlikely.  The lack of galaxies at the
redshifts of these systems leads us to suggest they are associated with the
supercluster filament (in which the galaxy is embedded).

A second case is that of Ton S180, a Sy 1.2 galaxy at a similar redshift to
(and possibly probing) the Pisces-Cetus supercluster.  Three
previously-unreported \lya\ systems are seen and matched with \ovi\
absorption at similar redshifts.  However, it is impossible to determine
the site of absorption, since it may be produced by the filament or by the
AGN itself.

The remaining five absorption systems are not associated with halos of
known galaxies.  They have no apparent galaxies within $200 h^{-1}$ kpc,
although this analysis is preliminary and based on shallow Digitized Sky
Survey images.  Further deep imaging around these sightlines is needed.
Recent statistical studies suggest that the proposed absorber-galaxy link
is valid only for strong \lya\ absorbers ($\ew \ge 400$ m\AA), and that
most local absorption systems are associated with filaments (Penton, Stocke
\& Shull 2002, 2004).

\section{Random, Unassociated \lya\ Clouds?}

It is possible that these absorbers are associated with neither filaments
nor intervening galaxies, and by chance appear at the redshifts where we
search.  Without worrying about their physical nature, only that they are
part of the ensemble of low-$z$ \lya\ systems well-studied in the
literature, we can derive the probably of detecting the number we have in
the regions we have searched.

For a velocity range $\pm \Delta v$, the expected number of 
absorbers is $\mu = 2 (\Delta v/c) (dN/dz)$, where $dN/dz$ is the frequency
of absorbers with redshift.  This number has been estimated to be $\sim 28$
for $\ew(\lya) \ge 240$ m\AA\ and $\sim 15$ for $\ew(\lya) \ge 360$ m\AA\
(e.g., Dobrzycki \etal\ 2002).  If we assume the absorbers are Poisson
distributed in redshift, and use $dN(\lya)/dz = 28$ for a mean occurence of
$\mu = 0.24$ per supercluster redshift range, then there is a 21\% chance
of finding at least one absorber within a given 2600 \kps\ section of a
pencil-beam spectrum.  Excluding Ton S180 completely, and calling the $\ew
< 240$ m\AA\ system toward H 1821+643 a non-detection, this leads to a
0.6\% probability of detecting \lya\ absorption in five of seven trials
(i.e., five of seven supercluster sightlines).  If we include Ton S180 and
H 1821+643, the probability drops to 0.01\%.

The probability that these are unassociated clouds is likely much lower
than this.  We see three distinct absorption systems in both the Aquarius B
and Pisces-Cetus superclusters, and as the $dN/dz$ studies would count
these as separate absorbers in their calculation, a more
accurate figure is found by including these as separate components.  This
adds three additional absorbers (since the LLS is likely associated with a
galaxy) and results in a probability of less than \eez{-5} that these are
chance detections.

\section{Conclusions and Future Work}

We have detected absorption due to possible filaments along seven of eight
supercluster sightlines.  From a statistical analysis, it is likely that
these absorbers are associated with the supercluster filaments, although it
is unclear at this point whether they are produced by hot diffuse material
in the filament or by some other mechanism.  Galaxies should trace the
filaments, and while we have argued against absorption due to galaxy halos,
it is possible we are seeing absorption from galactic winds or ejecta.  It
can be argued that such material is really part of the diffuse filament
gas, as it is gravitationally unbound from the galaxy and should eventually
thermalize with a hot ambient medium, if present.
Deep imaging of these sightlines will shed light on the presence of nearby
galaxies, and datasets such as the Sloan Digital Sky Survey will provide
redshifts for such systems and the clusters that lack redshifts.

The temperature of the absorbing gas is unknown at this time.  The
linewidths for most of the systems are consistent with temperatures up to
\eez{5}--\eez{6} K, although if multiple velocity components are blended to
produce the lines we see, this value will be much lower.  The lack of \ovi\
absorption remains puzzling, as it should be quite strong at the expected
temperatures.  Perhaps we are seeing cooler, denser streams of gas within a
more diffuse hot medium; modeling of the ionization state and metallicity is
needed before the observed $\lya/\ovi$ line ratios can be explained.

\begin{figure}
\epsfig{file=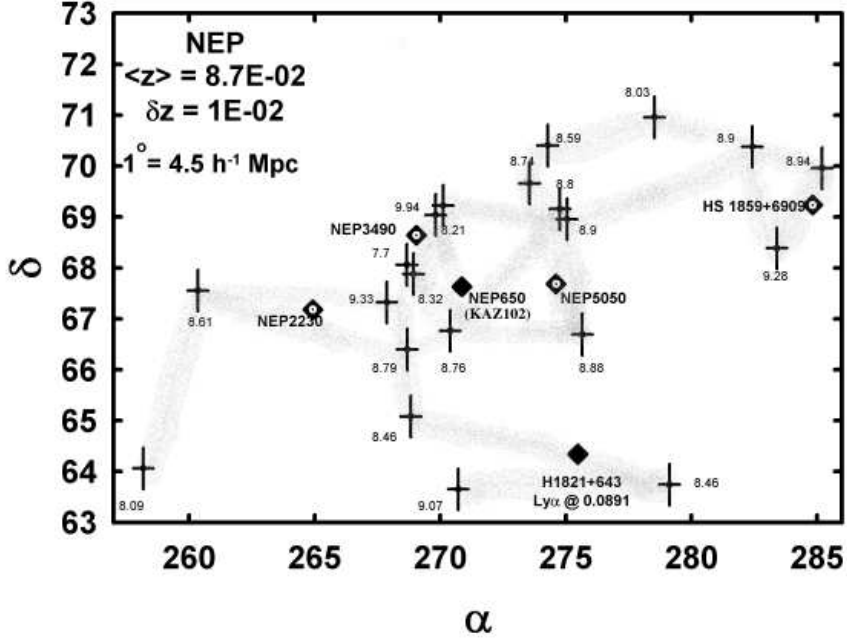,width=\linewidth}
\caption{Map of the North Ecliptic Pole supercluster, with the locations of
six background AGNs marked.  Two of the AGNs (solid diamonds) show \lya
absorption and are discussed in this work.  The remaining four (open
diamonds) have not been observed.
\label{fig:nep}}
\end{figure}

For the (uncertain) future of HST, there are several additional
observations that would enhance this study.  The North Ecliptic Pole
supercluster, shown as a map in Figure \ref{fig:nep}, lies in front of six
background AGNs.  Two of those have been previously observed and presented
here (H 1821+643 and KAZ 102), and both show \lya\ absorption.  Observations
of the other four would not only increase our target sample, but they would
provide an unprecedented study of the absorption properties across a
single supercluster.  Sightlines through other superclusters undoubtedly
exist in the HST archive as well, but with no comparable missions planned
for the foreseeable future, time is running out for new work in far-UV
spectroscopy.

\begin{acknowledgments}
We would like to thank the symposium organizing committee, especially Mario
Livio, for assembling such an interesting and valuable meeting.  Also, we
would like to thank B-G Andersson, Ken Sembach, Bart Wakker, Jimmy Irwin,
Chris Mullis, and Edward Lloyd-Davies for their advice and encouragement.
This work is supported by NASA grants NAG5-10765 and NAG5-10806.
\end{acknowledgments}

\end{document}